\documentclass[preprint]{aastex}        % but some stuff doesn't work!

\newcommand{\ergcm}{ergs~cm$^{-2}$~s$^{-1}~$}

\newcommand{\mdot}{$\dot{{m}}$ } 
  
\def\about{$\sim$} 

\def\erg/cm2sec{ergs~cm$^{-2}$~s$^{-1}$}   
\def\ergcm2{ergs~cm$^{-2}$}   
\def\cm2{cm$^2$} 

\received{}
\accepted{}
\begin{document}

\title{Timing Analysis of the Light Curve of the Dipping-Bursting X-ray Binary X1916-053} 

\author{Y. Chou, J. E. Grindlay, and P F. Bloser}
\affil{Harvard-Smithsonian Center for Astrophysics, \\
60 Garden Street, Cambridge, MA 02138\\
yichou@cfa.harvard.edu}

\authoremail{yichou@cfa.harvard.edu}
%\maketitle

\begin{abstract}

We present the timing analysis results for our observations of the 
x-ray dip source X1916-053 conducted with RXTE between February and 
October of 1996. Our goal was to finally measure the binary period -
as either  the x-ray dip period
or $\sim$1\% longer optical modulation period, thereby establishing if
the binary has a precessing disk (SU UMa model) or a third star
(triple model).  Combined with historical data (1979-96), the x-ray 
dip period is measured to be 
3000.6508 $\pm$ 0.0009 sec with a 2$\sigma$ upper limit 
$|\dot P| \leq 2.06 \times 10^{-11}$. From our 
quasi-simultaneous optical observations  
(May 14-23, 1996) and historical data (1987-96), we measure 
the optical modulation period to be 3027.5510 $\pm$ 0.0052 sec 
with a 2$\sigma$ upper limit $|\dot P| \leq 2.28 \times 10^{-10}$. 
The two periods are therefore each stable (over all recorded data) 
and require a $3.9087 \pm 0.0008$d beat 
period. This beat period, and several of its 
harmonics is also observed as variations in the dip shape. 
Phase modulation of x-ray dips, observed in a 10 consecutive day
observation, is highly correlated with the $\sim$3.9d dip 
shape modulation. The 1987-1996 optical observations show 
that the optical phase fluctuations are a factor of 3 larger 
than those in the x-ray. We discuss SU UMa vs. triple models 
to describe the X1916-053 light curve behavior and conclude that 
the x-ray dip period, with smaller phase jitter, is probably the binary period so that the 
required precession is most likely similar to that observed in SU UMa
and x-ray nova systems.  However the ``precession'' period stability 
and especially the fact that the times of x-ray bursts may partially
cluster to occur just after x-ray dips, continue to suggest that this system may be a hierarchical triple.

\end{abstract}

\section{INTRODUCTION}

The discovery of periodic ($\sim$50 min) dips in the x-ray flux 
from the moderately faint source 4U1915-05 (X1916-053) provided 
the first direct evidence for a binary periodicity in an x-ray burst 
source (Walter et al 1982, White and Swank 1982). Combined with 
the general success of the thermonuclear flash model for (Type I) 
x-ray bursts (e.g. Joss and Li 1980), this confirmed that 
bursters are indeed neutron stars accreting from binary companions. 
The x-ray dips are believed to be caused by partial 
occultation by vertical structure (or clouds) in the accretion
disk. The short dip period ($\sim$3000 sec), if identified as 
the binary period, provided the 
the first example of an ultra-compact x-ray binary with a very low mass
degenerate companion. 

The x-ray period has been reported with values from 2985 to 3015
sec (Walter et al. 1982; White and Swank 1982; Small et al. 1989;
Yoshida et al. 1995 and Church et al. 1997). Smale et
al. (1989) discovered a possible second candidate period of
2984.6 $\pm$ 6.8 sec from the Ginga 1987 observations.  In addition to
the primary dips, secondary dips with phase  
difference $\sim 180^{\circ}$ are 
sometimes observed.  The 
varieties of dip width, depth, phase, and association with the period
show the complex behavior of X1916-053.  In addition to the 3000sec
dip period, a 199d long-term x-ray modulation period was also reported by
Priedhorsky and Terrell (1984).

The optical counterpart of X1916-053 was discovered (Grindlay et al. 1988) as a V=21
mag optical star with a modulation period of 3027.5 sec, $\sim1\%$ longer than 
the X-ray dip period.  The difference between  the x-ray and optical
periods leads to the question of which represents the orbital period of the
system. The optical period was found to be stable over (at least) 
7 years (Callanan, Grindlay and Cool, 1995, hereafter CGC95).

In this paper, we briefly introduce SU UMa vs. triple system models
previously suggested (Grindlay, 1992) for
explaining the x-ray/optical period difference (section 2), describe
our 1996 RXTE observations  of X1916-053 (section 3) and
report (section 4) the timing analysis of the data, including the x-ray dip 
periodicity, $\sim$ 3.9d period dip shape modulation, the dip phase jitter 
phenomenon, and the long term x-ray dip period stability from the 
combination of the RXTE 
and previous x-ray observations (Einstein (1979-1981), 
EXOSAT (1983) and Ginga (1987-1990)).  
Analysis of our optical May 14-23, 1996, optical observations 
quasi-simultaneous with RXTE and historical (1987-96) 
optical data are used to derive the optical ephemeris and phase jitter 
for comparison with the x-ray values. We use the x-ray dip ephemeris, which 
is able to predict dip times within \about0.1 phase (300sec) back 
to 1979, to study the relation of the times of x-ray bursts to dips since 
a possible correlation was previously noted (Grindlay 1992). 
In section 5, we discuss SU UMa vs. triple models for the complex light curve behavior of 
X1916-053, concluding that either model would require interesting new
phenomena to explain the present data but that the triple model would
be required if the dip-burst correlation is real.

\section{SU UMa and Triple Models}

Grindlay (1989, 1992) pointed out that two models might account 
for the the beat behavior and the $\sim$1\% difference
of the optical and the x-ray periods.  The dual periods suggest either
a SU UMa type precession or a triple system.
SU UMa binaries are a subclass of dwarf novae, which during their superoutburst
state, display ``superhumps'' with periods a few
percent longer than the orbital period (see, e.g., Warner 1995).  Similar $\sim$1\% longer
periods are formed in the optical outbursts of x-ray transients.  For
both SU UMa dwarf novae and x-ray transients, the binary period is the
shorter of the two, and the longer period (which changes during
outburst decay) is the beat between the binary and a longer disk
precession period.  Applied to X1916-053, the shorter x-ray dip
period would be the binary period.  Whitehurst (1988) has
discovered by hydrodynamic simulation that an  accretion disk in a CV system
with a low-mass secondary is tidally unstable and leads to a
precessing 
eccentric disk.  Hirose and Osaki
(1990) have derived the dependence of the disk precession period on
binary mass ratio
which we apply to X1916-053 in section 5.1.

The triple model for X1916-053 was proposed by Grindlay et al. (1988)
and Grindlay (1989, 1992)
to account for the apparent stability of the
(longer) optical period and the reported 199d
long-term modulation (Priedhorsky and Terrell 1984).
Mazeh and Shaham (1979) showed that in a hierarchical triple system,
the eccentricity of the inner binary is modulated at a long-term
period $P_{long} = K
P^{2}_{out}/P_{in}$, where $P_{in}$ and 
$P_{out}$ refer to the binary period and the orbital period of the
third companion and K is a constant of order unity which depends on
mass ratios and relative inclinations.  Under the
triple model, the third companion period ($\sim$4d) in the X1916-053 system
should be the beat period of the x-ray
dip period and the optical period.  X-ray dips are due to mass
transfer ``surges'', from the change in inner binary eccentricity
induced by third star at the period given by the beat between the
outer and inner orbital periods.  If the optical period, the
longer one, is the binary period, then this triple companion is in a
retrograde orbit and the constant K=0.46 for X1916-053
system.  This model also predicts that the x-ray dips
should display phase glitches  (Bailyn 1987) and a timing correlation
between x-ray dips and x-ray bursts (Grindlay 1992, and see section 4.4).

\section{RXTE and Optical 1996 OBSERVATIONS}

\indent RXTE Observations of X1916-053 were made from 1996 February to 
October. A $\sim 10^4$ sec ($\sim$3 orbit) observation of X1916-053 was
conducted about once per
month except March and April.  Between May 14 to 23, the observations
were made about once per day.  The observation details and mean count
rates (outside dips and bursts) are given in
Table 1.  A total of 37 complete primary dips (hereafter, dips)
and 4 secondary dips with various depths and dip duration were
recorded.  A typical light curve is shown in Figure~\ref{tylc}.  All
observations except two contained dips.  No dip was observed in the May 5
observation due to Earth occultation.  However, for the May
15 data, two dips did not appear at the expected times (see
Figure~\ref{misdip}).

The data used for analysis is the PCA (PCU 0 and 1) 
Standard-2 data with time
resolution 16 sec.  We divide the data into 4 energy bands,
1.72-3.18keV, 3.18-5.01keV, 5.01-6.84keV and 6.84-19.84 keV. Since the
energy dependence is small in the dip centroid-timing analysis (see,
e.g. Church et al. (1998)) for analysis of spectral variation in dips
for X1916-053), we only 
present the analysis results for band 4 unless specified. A 
complete analysis of 
the persistent source (outside dips) spectra and variations of X1916-053 
for these RXTE observations is presented 
by Bloser et al (2000).

Optical observations of X1916-053 were carried out (by PFB) from 1996 May 14 to
23, with the Mt. Hopkins 48-inch Telescope with a broad-band B+V
filter. The exposure time of each optical image was 300 sec, but RXTE 
visibility constraints prevented the observations from being 
exactly simultaneous. The magnitude
of X1916-053 for each exposure was measured with DAOPHOT 
(Stetson, 1987) relative to 
the average magnitude of several nearby stars.  A reference 
star close ($\sim1$ arcmin)
to X1916-053 and within $\sim$1 mag was used to
measure the total measurement errors of the observation. Some data
are not included in the analysis because of the large fluctuation in
the comparison star.  Only five nights' observations (UT May
15, 16, 20, 21 and 23) have good light curves for analysis.  A typical
optical light curve is shown in Figure~\ref{tyoplc}.

\section{DATA ANALYSIS}

We first describe a period folding analysis for an initial
determination of the x-ray dip period.  This is then refined, and limits
on $\dot P$ are derived from the phase analysis. We then apply a similar 
analysis for the optical data and derive, similarly, a long-term
ephemeris. Finally, we use the dip 
ephemeris to examine the dip phase of x-ray bursts detected from X1916-053.

\subsection{X-ray Dip Periodicity Analysis}

\indent All data were initially corrected for arrival
times at the solar system barycenter.  We first carried out a $\chi^2$ analysis of the folding light
curves (32 bins) to search for the best
period of X1916-053 dips near 3000 sec for the entire 1996 RXTE data. 
The maximum $\chi^2$ is at 3000.6 sec as shown in
Figure~\ref{fesall}. Fitting the 3000.6 sec peak with a Gaussian gave
the best period of 3000.625 $\pm$ 0.192 sec. The side bands are mainly the 
results of the beat period of observation gaps.

Performing the same $\chi^2$ analysis on the ten consecutive
observations in May, we find that there is a family of peaks, in
addition to the 3000 sec peak,
including a $3026.23\pm 3.23$ sec peak near the 3027 sec optical variation
period.  The centroid of these
peaks associated with the 3000 sec period implies that there is a modulation
with a fundamental period of $\sim$3.9 days, as previously noticed
(Grindlay 1992), in the X1916-053 x-ray
light curves (see Figure~\ref{fesmay}).  If we take the modulation
period to be the 3.9087d beat period of 
the optical period (3027.5510 sec; cf. section 4.3) and
the x-ray dip period (3000.6508 sec; cf. section 4.2) respectively, 
we may then define the x-ray side band periods as 

$${1 \over {P_n}^{side}} = {1 \over {P_x}} - {n \over {P_m}}\eqno(1)$$ 
where $P_x$=3000.6508 sec, $P_m$=3.9087d, n= $\pm$1,$\pm$2
..... and $P_m$/n is the nth harmonic of the modulation period.
Table 2 shows the expected side band periods and the measured
side band periods.

To further study the origin of the $\sim$ 3.9d beat period, we folded the
light curves with the folding period of 3000.625 sec for
each of ten consecutive days. The daily folded light curves are
grouped into 4 day groups in Figure~\ref{tenfef} to demonstrate the light curve
variations.  The $\sim$3.9d period
is evident as variations in dip shape, especially duration.  In Figure~\ref{tenfef}, the first
column (May 14, 18 and 22) has the largest dip widths, and the
width narrowed in the following 3 days (except May
15 when, surprisingly, no dip was observed). The second 
row (May 18 to 21) of the plot shows a complete period of 
the width variations.  On the last day
of each cycle (May 17 and 21), the secondary dips appear so that these are
also tied to the $\sim$3.9d clock.  The dip depths also have a variation
period of $\sim$3.9d (see Figure~\ref{tenfef}) but the change is not as 
significant as the variation in the dip widths.

Schmidtke (1988) reported two possible optical periods of
2924.76 $\pm$ 2.1 sec and 3027.48 $\pm$ 2.2 sec in observations
from the CTIO 1.5m telescope on UT 1987 May 3, 4, 5 and 8. 
Table 2 suggests that the 2924.76sec period is probably an observation of the 
harmonic index -3 side band.
Grindlay (1989 and 1992) reported that the x-ray period was found in
the optical light curves also. Together with the side bands and dip
width-intensity modulation, we now find 
in the RXTE x-ray light curves, the reality of a stable  
$\sim$ 3.9d period in the X1916-053 
system is now established.

\subsection{X-ray Dip Phase Analysis}

We use the phase of the dip center time folded
by a trial ephemeris to refine the dip period, measure the phase variations, update the ephemeris and derive the
period derivative. The dip center time is 
obtained from fitting a quadratic curve around the minimum of the
dip and then folding it with a trial period and an 
arbitrary but fixed start epoch as the phase zero.

The phase analysis can be used as an independent means to 
determine the best fit period in addition to the 
folding light curve analysis in section 4.1.
The first trial period may be taken arbitrarily but must be close to the
true period so that there is no cycle count ambiguity for neighboring
observations. We choose the period from a $\chi^2$ analysis of all 1996
light curves to be the initial trial period.  The slope of the linear
fit of the phase $\Phi$ and dip center time for the data obtained over
time interval t gives the correction of the period as

$${\Delta \Phi \over {t}} = {\Delta P \over {P^{2}_{fold}}}\eqno(2)$$

\noindent We repeat the correction process until the phase drift
rate (i.e. the slope) is much smaller than the error of the slope of the
linear fit. Similar to the period correction, the error of the period
is derived from the error of the fit slope.  For the RXTE 1996 February to
October data set, the best period from a linear fit is 3000.625 $\pm$ 0.02 sec 
(see Figure~\ref{phase96}) and the RXTE x-ray dip ephemeris can be written as

$$T_{RXTE}=MJD50123.00944 \pm 1.4 \times 10^{-4} +(3000.625\pm
0.02)/86400 \times N\eqno(3)$$

To further improve the accuracy of the x-ray dip period as well as to
study the long term stability of X1916-053 period, we analyzed
historical data. First, we folded the dip 
center times of Ginga '87, '88, and '90 observations 
(cf. Yoshida et al 1995) with eq.(3). Together with 
the RXTE observations, we found
a phase drift with a constant rate, which implies that a constant
period can phase all the 1987-1996 data.  The period error of 0.02 sec
in eq.(3) is equivalent to a phase error of 0.07 per  year.  The
probability of cycle count ambiguity can be written as (assuming the
distribution is Gaussian)

$$Prob = {\int_{-0.5}^{0.5}} [1-{{exp({-{{\Phi^2} \over {2 \sigma^2}}})
/  {{\sum_{n=-\infty}^{\infty}}{exp(-{{{{(\Phi+n)}^2} \over
{2 \sigma^2}})}}}}}]  d \Phi\eqno(4)$$

\noindent The 0.07 phase error per year gives the
probabilities of cycle count ambiguities 0.007, 0.027 and 0.199 for the
'87 to '88, '88 to '90 and '90 to '96 data, respectively.

We divide the RXTE observations into two parts, the first and the
second halves of 1996.  Together with the Ginga '87, '88 and '90 observations, 
we have a total of  5 data sets (all, of course, 
corrected to the barycenter).  
We take the average phase to be the mean
phase of each dataset and assign the mean phase fluctuation to
be the error of the phase.  The linear fit for the Ginga-RXTE data 
refines the dip period to be P = 3000.6508$\pm$0.0021 sec. 

Next, we add the Einstein '79-'81 data (Walter et 
al. (1982) and White and Swank (1982)) 
and EXOSAT '83 data (Smale et al. (1988)), again all barycenter
corrected, and
fold with the 3000.6508 sec period and the ephemeris of eq.(3). 
In the EXOSAT data sets, folding 
the dip center times with the ephemeris above to the 1985 May and October
data  (3 dips) yields phase of $0.48 \pm 0.02$.  Thus these are secondary
dips and are excluded from further  analysis. We assign a $\pm 0.05$
phase error from the phase jitter range to each dataset.  The
x-ray ephemeris can phase all of the 1979-1996 dataset within a phase
error of $\pm 0.07$ with no clear systematic derivative.  
The linear fit result 
shows that the phase drift rate is $2.19 \times 10^{-6}$ per day,
which is smaller than the phase drift error $8.37 \times 10^{-6}$ 
per day (1$\sigma$). Hence, no period correction is necessary for the x-ray
ephemeris and the best fit x-ray dip ephemeris (hereafter, the x-ray ephemeris) is:

$$T_{dipcenter}=MJD50123.00944 \pm 1.4 \times 10^{-4} +(3000.6508
\pm 0.0009)/86400 \times N\eqno(5)$$
\noindent The quadratic fit of $\Phi$ vs. t gives the period derivative
because

$$P(t) \approx P_0+\dot {P} t = P_{fold} + \Delta P + \dot {P} t\eqno(6)$$

\noindent Thus,  the phase can be written as

$$\Phi \approx \Phi_0 + {{\Delta P} \over {(P_{fold})^2}} t + {1 \over
2}{{\dot P} \over {({P_{fold}})^2}} t^2\eqno(7)$$

\noindent where $P_{fold}$ is the folding period, $\Delta P=P_0-P_{fold}$
and $\Phi_0$ and $P_0$ are constants. 
The X1916-053 x-ray dip period derivative from a quadratic fit for all
1979-1996 data sets is

$${\dot{P} \over {P}} = (6.50 \pm 10.86) \times 10^{-8} yr^{-1}\eqno(8)$$

\noindent The minimum $\chi
^{2}$ fit gives $\chi ^{2}$ values of 2.00 and 1.64 and a reduced ${\chi
^{2}}_{\nu}$ of 0.286 and 0.274 for linear and quadratic fits, respectively.  
From the one sided F-test, we obtain $F=\Delta \chi^{2} / 
{{{\chi}^{2}_{\nu,quad}}} =
(2.0-1.64)/0.274 =1.31$ in the comparison of linear and quadratic fits.
Thus the quadratic fit, 
shown in Figure 7, is not significantly better than the
linear fit.  We derive a 2$\sigma$ (90\%) confidence
level upper limit for any change in the x-ray dip period as 
$|\dot P_{x-ray}| \leq 2.06 \times 10^{-11}$. 

When the dip centers of the RXTE 1996 May
data are folded by the x-ray ephemeris (shown in Figure~\ref{jitter}), a phase
systematic variation  is clearly seen on the time vs. phase plot.  In the RXTE 
1996 May
observations, the phases form two groups. The dips from May 16, 17 and 21
have phases larger than 0 and all the others (except May 15, with
missing dips) smaller than 0. The range of the phase jitter is about
$\pm$ 0.05 in phase.  The phase jitter seems to have some 
correlation with the 3.9d
period. We fit a 50\% duty cycle square wave with fixed 
3.9087d period (the beat period
of x-ray and optical periods). The  minimum $\chi
^{2}$ fit result is shown as Figure~\ref{jitter} and the reduced
$\chi^2$ is 0.45 for 19 degrees of freedom. 

Yoshida (1993) and Yoshida et al.(1995) fit the dip phase variation
with a sinusoidal period $6.5 \pm 1.1$d from Ginga 1990 September
observations, so we did a similar analysis for the RXTE 1996 May observations. 
There are two (local) $\chi^2$ minima between the periods of 3.0d to 7.0d,
$4.85 \pm 0.12$d and $3.76 \pm 0.10$d, with amplitudes of $\sim 0.06$
phase (or $\sim$3 min) and reduced $\chi^2$s 0.49 and 0.60, respectively (17 
degrees of freedom).
Both fits are good.  However, although the $3.76 \pm 0.10$ d  period
result has slightly
larger reduced $\chi^2$, it is consistent with the 3.9d phase jitter, or the
dip width modulation  period. 
The sinusoidal fit with fixed period of 3.9087d gives a reduced
$\chi^2$ of 0.69 (18 degrees of freedom).  All the fit results are
shown in Figure~\ref{jitter}. Thus the 3.9d period found for the dip 
phase modulation with RXTE is also within $\sim1\sigma$ 
of 0.5 the Ginga period (since 3.9 $\sim$ 0.5(6.5 + 1.1)).

\subsection{Analysis of optical light curves}

The stability of the optical modulation period is critical for us to decide which
model best describes the X1916-053 system.  The optical period
was confirmed by CGC95 to be stable over 7 years and the times of minima
can be described by a simple ephemeris.  We have now found the x-ray period to 
be stable for  $\sim$ 17 years (see section
4.2).  Both periods have comparable long-term stability with significantly
different period values.  However, in the long-term stability analysis
in Grindlay (1989), CGC95 and section 4 of this
paper, the minimum phases are averaged over several days to several
months. Those long-term analyses ignore the 
short-term phase fluctuations (say, several
days). Figure~\ref{phstat} shows the X1916-053 phase distribution of
both x-ray and optical 1987-1996 observations folded by the
corresponding ephemerides for x-ray (eq. (5)) and optical (from CGC95) data 
respectively.  
The mean phase of the x-ray dips is 0.00098 with 1 $\sigma$
fluctuation of 0.06, and none of the 101 dips exceeds phase $\pm$
0.2.  On the other hand, the optical minima have a mean phase of 0.0573
and 1 $\sigma$ fluctuation of 0.151 using the CGC95 optical ephemeris.
This shows that the optical minima have larger 
fluctuations than the x-ray dips.  
The $\sim$ 0.05 offset in the
optical minima phases is correction of the phase zero epoch
given by CGC95 that is required to fit the 1996 optical data.

With the new dataset of 1996 observations, the long-term stability
of X1916-053 optical modulation can be tested again.  We folded the
light curves by the optical ephemeris from CGC95 for each 
night's observation to 
obtain the minimum
phase of the night.  The same method described in section 4.2 is applied
to get the average minimum phases and the average observation times for each
dataset.  We assigned 0.15 phase error obtained from the mean
fluctuations in phase to each dataset. The linear fit shows that the
optical ephemeris results in a phase drift of $5.07 \times 10^{-6}$ per
day with 1$\sigma$ error of $4.93 \times 10^{-5}$ per day. Correction of the 
period from CGC95 is
unnecessary because the phase drift is less than the error.  The linear fit
demonstrates that the ephemeris in CGC95 is still good but
we have obtained a
smaller error of the period at $5.23 \times 10^{-3}$ sec (= $6.06
\times 10^{-8}$d) than the one in CGC95 (=$8 \times
10^{-8}d$). Incorporating the 0.05 phase offset, we derive the final X1916-053
optical ephemeris

$$T_{optmin}=HJD2444900.0046 \pm 0.003 +(3027.5510 \pm
0.0052)/86400 \times N\eqno(9)$$

\noindent From the quadratic fit, we obtain a period derivative of

$${\dot{P} \over {P}} = (8.9 \pm 11.88) \times 10^{-7} yr^{-1}\eqno(10)$$

\noindent which could be $\sim$14 times larger than that from the x-ray
dips but again is not a significant detection of $\dot{P}$.  Both linear and
quadratic fits yielded good fits with small reduced $\chi^{2}$ values
of 0.169 and 0.106 respectively.  The F value of the one-side F-test is 6.3
for comparison of the linear fit and the quadratic fit. 
This value implies that 
the quadratic fit, as shown in Figure ~\ref{qfitop}., is better 
than the linear fit at the $\sim$95\% confidence level. The 2$\sigma$ (90\%) 
confidence level upper limit for the change in optical period is  
$|\dot P_{opt}| \leq 2.28 \times 10^{-10}$

\subsection{Analysis for Triple Model}

Grindlay et al. (1988) and Grindlay(1989,1992) proposed a 
triple model in which the third star period was
related to the beat period of optical and x-ray periods ($\sim$ 3.9d).   
In a hierarchical triple, the predicted (Bailyn 1987) 
phase of the minima of the inner binary separation, and thus (potentially) the 
mass transfer rate \mdot  would be modulated at the beat between the
outer ($\sim3.9d$) and inner binary (3000 sec) periods.  The x-ray dips 
are due to these \mdot modulations which ``puff up'' the 
accretion ring (where the accretion stream strikes the inner disk). So
that dips do not occur at a preferential inner binary phase.

Bailyn (1987) shows that the phase of the inner binary separation
should ``glitch'' twice per outer orbit, so we expect corresponding
phase jumps in x-ray dip times.  Indeed the possible square wave 
modulation of x-ray dip phase (Figure~\ref{jitter}) is suggestive. 
However, the observed dip phase modulation is about a 
constant x-ray dip period, whereas the phase glitches 
expected (Bailyn 1987) in the inner binary separation and 
thus dip times are additive: they effectively delay the 
times of dips so that over the outer orbit period there are 
as many total minima (dips) as there are inner binary 
orbits. Thus the x-ray dip phase should not depart 
from the binary phase by more 
than \about0.5 (depending on the ``glitch spacing ratio'', 
in turn dependent on outer vs. inner binary mass ratio). 
On the other hand, if the x-ray and optical clocks 
are independent, their relative phase offsets (on either the x-ray 
or optical ephemeris) will be randomly spaced. Thus, the free running x-ray 
and optical phases folded by the
``wrong'' period (e.g., folding the optical minima with x-ray
ephemeris or folding the x-ray dips with optical ephemeris), will make
the phase offsets drift linearly.  If the observation times are random,
the phase difference will be random and give rise to a uniform (with certain
statistical fluctuation) dip (or minimum) phase difference distribution. Grindlay
(1992) found marginal evidence from the 1990 Ginga and optical data for correlated optical minima vs. x-ray 
dips.

We therefore analyzed the phase difference of the simultaneously-observed
optical minima and x-ray dips.  A total of 12 pairs of (x-ray and optical)
dips/minima in the '87-'96 datasets 
(9 primary and 3 secondary dips/minima) have dip-center/minimum time differences less
than 1500 sec and can thus be compared in phase for the same 
binary orbit.  Figure~\ref{dphstat} shows the phase 
difference distribution of these quasi-simultaneous dips/minima.
Doing a simple $\chi^{2}$ analysis 
for phase differences gives a reduced $\chi^{2}$ of \about1 when folded
by either the  x-ray or optical ephemeris.  
Both phase differences are thus consistent with a uniform distribution. 
Interestingly, though, the 3 secondary x-ray dips are  
clustered in a somewhat narrow phase range (-0.1 --- +0.3); clearly 
a larger sample is needed.

The triple model makes one other prediction (Grindlay 1992) : that there may be 
an association between x-ray dips and x-ray bursts. This is 
because each accretion surge (resulting in a dip), occurring 
at the dip period of successive minimum separation times, 
enhances the probability of a nuclear flash at times just 
following. The dip, due to obscuration by enhanced material 
and clumps in the accretion ring, precedes the flash by the 
timescale required for it to accrete from the ring onto the NS. 
This must be small, since the mean height of the ring must decay 
promptly (in the triple model) for the dips to have
their observed typical duration 
of \about0.1 - 0.2 phase. This then suggests the ring empties 
out, and accretes, via an ADAF flow (Narayan and Yi 1995) since the
viscous timescale for accretion in a thin disk from the ring at R \about 10$^8$ cm 
in this ultra-compact LMXB would still be $\geq$3000 sec, or 
at least a full binary period. Therefore if accretion onto the NS 
is enhanced by the rapid decay of the ring, perhaps 
through quasi-spherical accretion, 
the probability of the accreted material 
reaching the critical density to flash is expected to be greatest 
for times just after a dip.

In Figure~\ref{burststat} we show 
the phase distribution of 41 x-ray bursts as recorded by 
OSO-8, EXOSAT, Ginga and RXTE when folded on our final 
x-ray dip ephemeris (with phase 0 the phase of 
primary dips). One third of the bursts concentrate in phase 0-0.2 (after
primary dip).  The distribution has
a $\chi^{2}$ probability of 0.006 of being random (flat).  The
binomial probability of detecting 17 bursts in the ``interesting''
phase interval $\sim$0 --- 0.2 (required to match typical dip
durations (cf. Figure~\ref{tenfef})
and thus enhanced accretion time) relative to the uniform distribution (8
bursts) is 0.016.  No such effect is seen if the burst times are
folded on the optical period (dashed line in Figure~\ref{burststat}).

\section{DISCUSSION}

Table 3 shows all the x-ray dip
periods reported since 1982.  The side bands from  the RXTE
analysis of X1916-053 may explain the many different reports of the x-ray dip
periodicity of X1916-053 .  Owing to the insufficient time span of
the observations, both uncertainties of the 3000 sec period peak and its first
neighbor side bands are large such that the peaks merge to
affect the centroid of the 3000 sec peak.  The centroid shifts left or
right depending on which (first left or right) side band is more
significant.

The analysis in this
paper suggests that there is a $\sim$3.9d period modulation in the dip
width and depth, and the phases may also have a $\sim$3.9d period jitter
with an amplitude of 0.1 (peak-to-peak).  The 6.5$\pm$1.1d period for
dip phase reported by Yoshida et al. (1995) is probably the
sub-harmonic of 3.9d period.  The optical modulation also has a $\sim$4d period
in amplitude, and possibly
phase modulation (Grindlay 1992), which is consistent with our RXTE x-ray
observations.

The fact that both the x-ray and the optical light curves 
have the $\sim$ 3.9d beat period
implies that there are 3 periodicities in the X1916-053 system, with periods
of approximately 3000 sec, 3027 sec and $\sim$3.9d. Of these three periodicities, at 
most two of them
are independent because of the beat relation among them. 
One of the two
short (3000 sec and 3027 sec) periods should represent the binary orbital
period and it should be the more stable one.  Both the optical and the
x-ray period have
long-term stabilities over the total historical data span of $\sim$9 years and
$\sim$17 years respectively.  The period derivative of the 
optical modulation could be
$\sim$14 times larger than that for the x-ray dips, although both
$\dot{P}$ values are statistically
insignificant.  The phase jitter statistics show that the x-ray dips have
smaller fluctuations than the optical minima.  Thus
the x-ray dip period may be the better candidate for the orbital
period, and thus X1916-053 contains a precessing disk as in SU UMa
systems.  However, the larger optical phase jitter could simply be due
to the shallower modulation in the optical vs. x-ray (compare
Figure~\ref{tyoplc} and Figure~\ref{tenfef}).

As outlined in section 2, SU UMa and triple models have been proposed
to explain the $\sim$1\% difference of x-ray dip and optical variation
periods.  We consider additional implications of each model. 

\subsection{SU UMa Model Implications}

Our result that the x-ray dips have less phase jitter, and 
are at least (if not more) stable over the historical database, 
provides strong support for the SU UMa model. 
Applying the SU UMa model to the X1916 system, the
3000 sec x-ray dip period is the orbital period which is due to the
shadow of accretion stream which crosses our line of sight $\sim$0.1
in binary phase before the secondary star. 
The 3027 sec optical period (the
``superhump'' period) is due to the beat with the 3.9d 
precession period of accretion disk.  The asymmetric disk and its $\sim$3.9d
retrograde precession that changes the vertical 
structure (cloudlets) at the outer edge    
with $\sim$3.9d period, provide an explanation for the 3.9d modulation
of x-ray dip shape and phase.

Hirose and Osaki (1990) performed hydrodynamic simulations of 
accretion disks for the superhump phenomenon in SU UMa stars.  They
showed that the disk precession angular frequency can be written as

$$  {{{\omega}_p} \over{{\omega}_{orb}}} = a(r) {q \over
{(1+q)^{1/2}}}\eqno(12)$$

\noindent where
$$a(r)={{1\over 4} {1\over {r^{1/2}}} {d\over {dr}}  {({r^2}
{{db^{(0)}_{1/2}(r)} \over {dr}})}},\eqno(13)$$

\noindent q is the mass ratio, r is the ratio of disk radius and
binary separation and ${b^{(0)}_{1/2}(r)}$ is the Laplace coefficient
of order 0 in celestial mechanics (see Brouwer and Clemence 1961, equation 42,
Chapter 15). Hirose and Osaki also demonstrated that the tidal instability of
accretion disks is caused by the parametric resonance between particle
orbits in the disk and the orbiting secondary star with 1:3 period
ratio.  Then, from Kepler's third law, $r=(1/3)^{2/3}=0.48$ for the
tidal instability radius.  From the Laplace
coefficients in Brouwer and Clemence (1961), $a(r=0.48)=0.41$.  Applying this model to
the X1916-053 system, for which $w_p / w_{orb}$= 0.0089, the mass ratio is 
q = 0.022,
or the mass of the secondary star $M_2$=0.03$M_{\odot}$ for the assumed 1.4
$M_{\odot}$ neutron star. This compares favorably with the value
$M_2$=0.016$M_{\odot}$ derived from the requirement that Roche
lobe-filling secondary has $R_{WD}\sim
R_{RL}$.  Thus the secondary is indeed a helium
white dwarf, as expected from its binary period.

The analysis in this paper indicates that the x-ray
dip period is more likely to be the true orbital period, which
supports the SU UMa model.  The only problem may be the stability of the
superhump period. The optical period derivative  obtained by the 1987-1996 
dataset is
$|\dot{P}|<2.28 \times 10^{-10}$, whereas the SU UMa stars typically have
$\dot{P} \sim -(3-10) \times 10^{-5}$ (Patterson et al. 1993a).
However, Patterson et al.(1993b) showed that the range of superhump period 
derivatives is
 $|\dot{P}| \sim 10^{-3} - 10^{-9}$ (for V795 Her, $|\dot{P}| <1.7
\times 10^{-9}$, see Zhang et al. 1991).  
 
AM CVn(=HZ 29), a binary with a white dwarf primary star and low mass
helium white dwarf secondary star, is a compact 
binary system which has behavior
similar to X1916-053.  It has two\footnote{In fact, AM CVn also has a 1011 sec
period which perhaps comes from the beat period of the 1028 sec period and
a 16.96 hr nodal regression period
(Patterson et al. 1993c, Skillman et al. 1999).} 
periods, 1028 sec and 1050 sec.  Patterson
et al. (1993c) discovered the 13.38 hr period, the beat period of the
1028 sec and 1050, in the helium absorption-line profile and suggested
that its accretion disk is elliptical and slowly precesses around the
white dwarf.  Provencal et al. (1995) and Solheim et al. (1998)
analyzed the Whole Earth Telescope (WET) data and discovered
the 1051 sec period is very stable 
with period derivative of $\dot P(1051s) = +(1.71
\pm 0.04) \times 10^{-11}$($\sim$ 10 years baseline)\footnote { The
1051 sec period wanders with $\dot{P}= \pm 2 \times 10^{-8}$ (time
scale of 4-12 months) but the
mean period has not wandered more than 0.1 sec in 35 years, which leads to
$|\dot{P}| < 10^{-10}$ (Skillman et al. 1999).}. 
They concluded that 1051 sec is the
orbital period and 1028 sec is the superhump period which would be
opposite to the normal superhump system and not supported by the tidal
instability model.  On the other
hand, Skillman et al. (1999) found that the period derivative of
1028 sec period is even more stable with $|\dot P(1028s)|< 2 \times
10^{-12}$ (over 7 years baseline).
Thus, they concluded the shorter period is the orbital
period and the longer one is the superhump period, which 
is comparable with our
results for X1916-053.  Both systems thus have very stable 
long-term superhump periods 
(${\dot P}_{sh}$(AM CVn)=$+1.7 \times 10^{-11}$ 
and ${\dot P}_{sh}$(X1916-053) 
$= P_{opt} \times {{\dot{P}}_{opt}} /{P_{opt}} \sim +8.6 \times
10^{-11}$, 
$\sim10^6$ times smaller than ones from typical
superhumps of dwarf novae). Therefore, at least some of the ``superhump CVs'' 
(e.g. AM CVn, V795 Her\footnote{These
are not true SU UMa (superhump) systems since neither system has been
observed to have outbursts.})
have very stable superhump periods comparable
to X1916-053.

The very stable superhump period implies a corresponding stable mean
mass transfer rate, and thus x-ray luminosity.  This is because
conservative mass transfer (assumed) at rate ${\dot{M}}_2$ implies $\dot{P}$/$P$
=${\dot{M}}_2$/$M_2$ (van der Klis et al. 1993) where $M_2$ is the
secondary star mass.  Our limit (eq. (8)) implies that ${\dot{M}}_2 \leq
 6.4 \times 10^{-9} M_\odot yr^{-1}$, or $L_x \sim 0.15 \dot{M}_2c^2 \leq 5.4
\times 10^{37}$ erg/sec, which is a factor of $\sim$10 larger than the typical system
luminosity measured from our RXTE data (Bloser et al. 2000).  Thus, measurement of $\dot{P}/P\sim 10^{-10}$, combined
with measured variations in $L_x$ on the $\sim$190d period (see
below), might allow a conclusive test of the SU UMa model.

\subsection{Triple Model Implications}

Although the SU UMa (or
superhump) model is the conservative choice, we
cannot exclude the triple model (Grindlay et al 1988, 
Grindlay 1989) since it could also 
explain the reported (Priedhorsky and Terrell 1984; hereafter PT84) 199d
long-term modulation of X1916-053 (cf. section 2). 

We have searched for the $\sim$199d period by an epoch-folding
analysis.  We find comparably significant peaks in a $\chi^2$
vs. period plot at $\sim$160d and 190d but not at the 198.6$\pm$1.72d value
found by PT84.  Longer epoch
analysis for the 199d period is needed, since the $\sim$4 yr RXTE/ASM
data span is short enough that a ``local'' period departure of the
Mazeh-Shaham period is possible (Mazeh, private communication). Extended x-ray and
optical coverage is also needed to reduce the limits on $\dot{P}$ for both 
the x-ray dip and optical periods. Although the predicted (Bailyn
1987) phase glitches for the x-ray dips, and the implied 
phase clustering of x-ray dips vs. optical (binary) minima, are not 
found in our phase analysis (except possibly for secondary dips), 
more simultaneous x-ray and optical observations 
are needed. 

The strongest remaining evidence for the triple model is the 
surprising clustering of x-ray burst times vs. the times of 
x-ray dips (cf. Figure ~\ref{burststat}). This would be expected under the 
triple model if the enhanced matter in the accretion ring can 
be accreted very rapidly onto the NS: for bursts to occur within the
$\sim$0.2 phase bin following a dip, the accretion time must be
$\leq$600 sec. This itself would be 
almost as interesting as the retrograde triple system nature (and its 
origin; possibly in a globular cluster), since such a  
surprisingly short accretion timescales implies accretion via an
ADAF-like quasi-spherical flow, rather than a dense accretion disk,
from the ring down to the NS.  Although the dip vs. 
burst association appears to be significant, more data are 
(as usual) needed. Quasi stable burst repetition times 
have been seen in one other source, GS1826-238 (Cocchi et al 2000), 
but without the remarkable stability apparent in X1916-053. \\

If the bursts are indeed correlated with dips (i.e. follow by
$\sim$0.0-0.2 phase), then under the simple binary and SU UMa model this would imply bursts
are preferentially detected when the $\sim$0.03$M_{\odot}$ WD
secondary is approximately aligned between us and the NS (i.e. when the NS
would be eclipsed if the system  were at higher inclination).  We
consider two very unlikely possibilities: lensing and beaming.  The WD-NS
alignment might suggest gravitational lensing or
focusing small bursts on the NS otherwise not detectable.  However, this would not
preferentially amplify bursts since the accretion luminosity would be
similarly lensed.  The near-eclipse WD-NS alignment could imply
burst emission is somehow beamed, perhaps by the WD-NS accretion stream-tube
which might confine the burst at its footprint on NS by magnetic
pressure.  The lensing or beaming scenarios are both so implausible
that confirmation of the burst-dip association would (virtually)
require the triple model.

A final test of the triple vs SU UMa models could be made by
irrefutably establishing the burst correlations (triple model) or
spectroscopically showing that the dip period is the binary period (SU
UMa model).

\acknowledgments

The authors thank D. Barret, J. Halpern and T. Mazeh for discussions, K. Yoshida for
provision of Ginga results and the HEASARC, from which archival OSO-8,
EXOSAT and Ginga light curves were obtained for dips and bursts.  This
work was supported in part by NASA grant NAG5-3298.\\

\clearpage

%figcaption

\figcaption[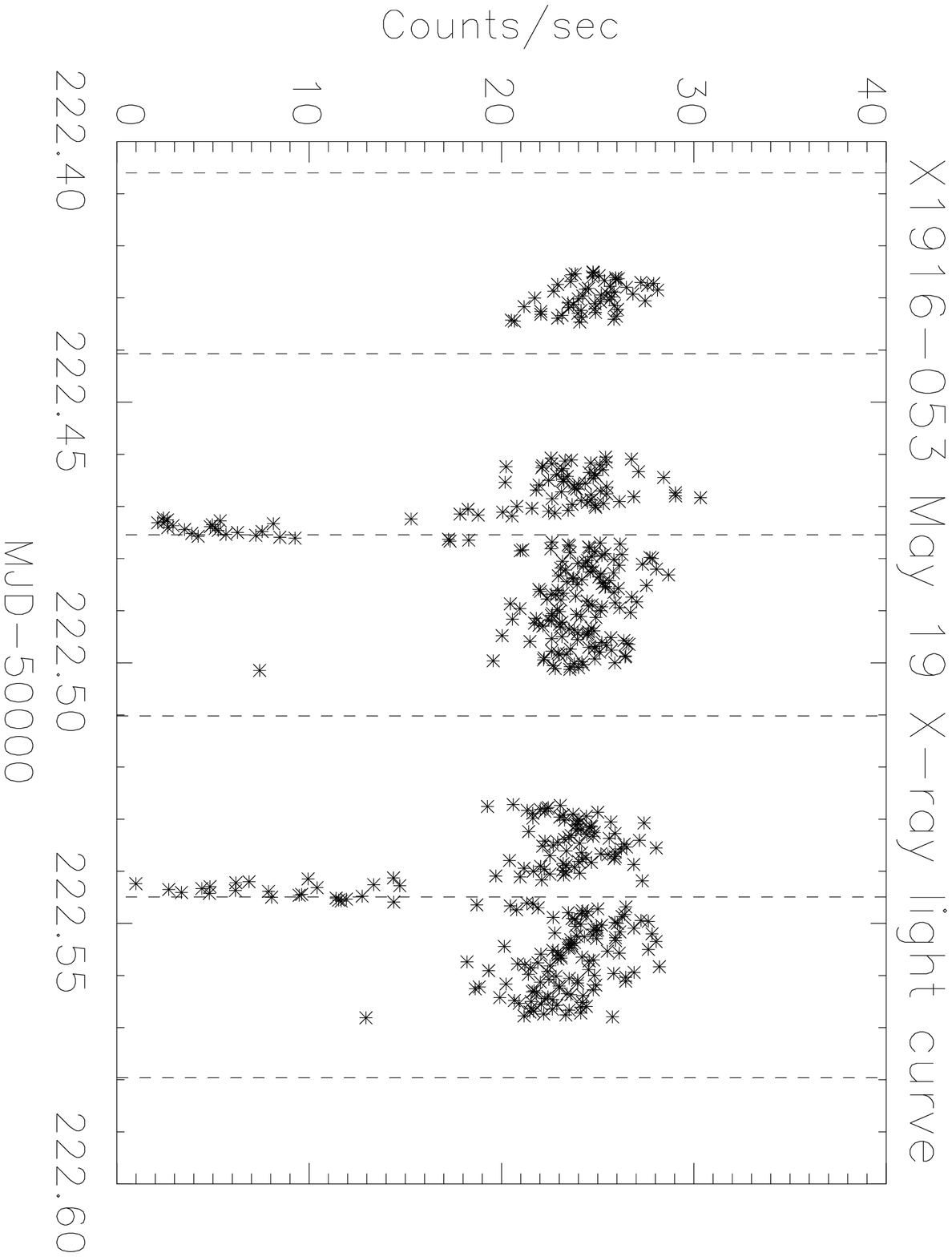]{A typical X1916-053 dipping light curve.  
The dashed lines are the expected dip center times for the ephemeris derived 
in section 4.2.
\label{tylc}}

\figcaption[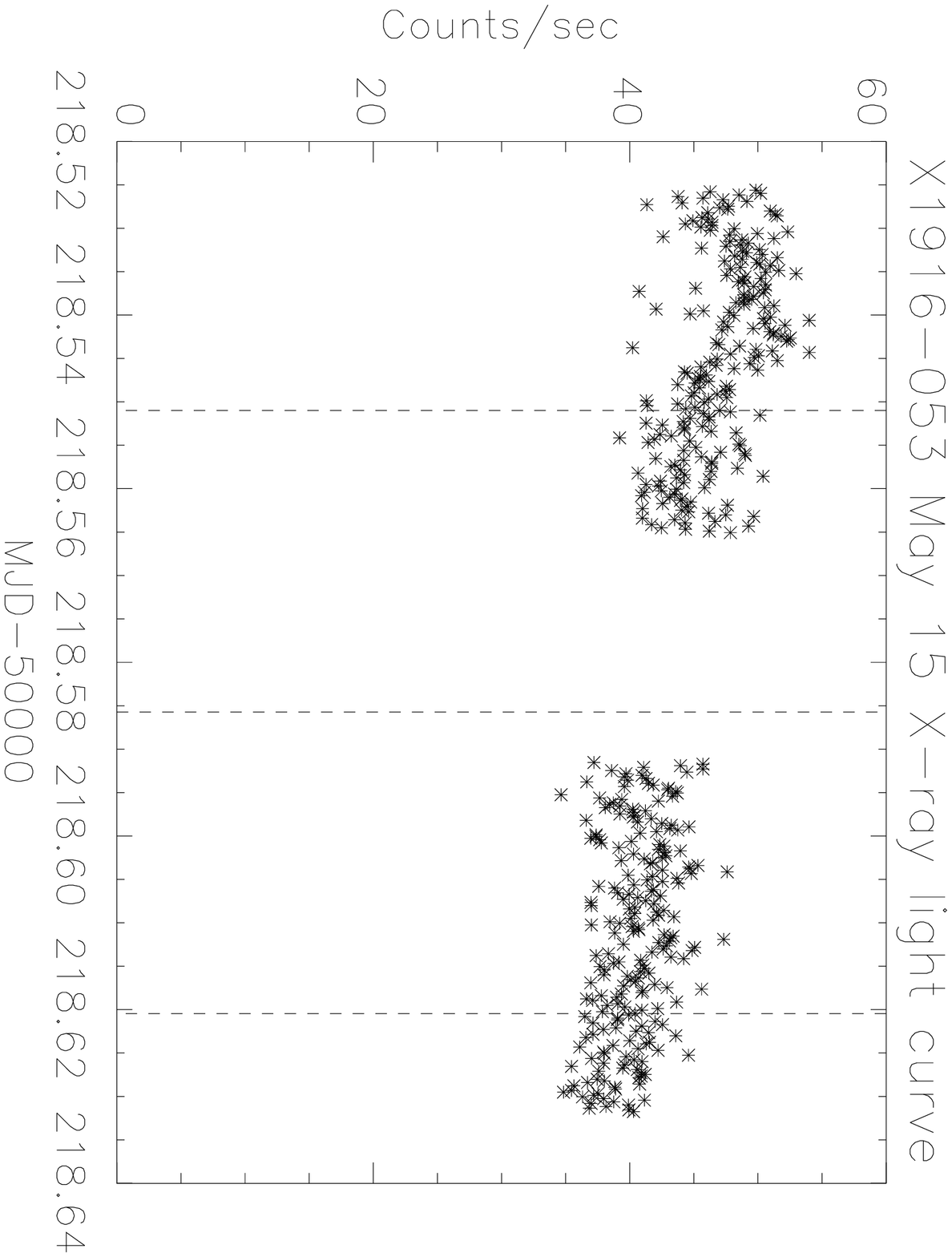]{May 15th observation.  No dip is 
observed at the expected dip
center times (the dashed lines).
\label{misdip}}

\figcaption[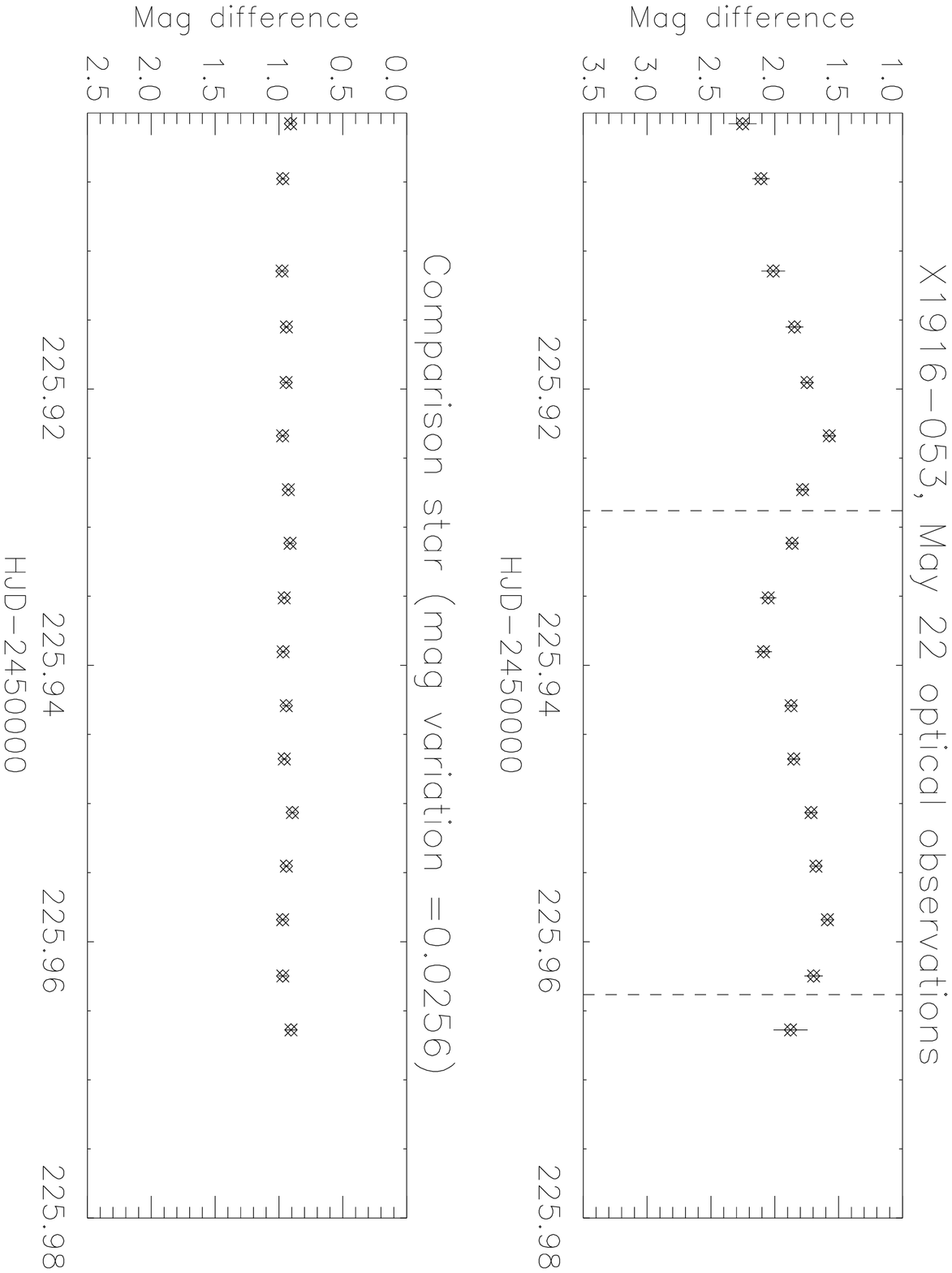]{A typical optical
(B + V band) light curve observed on UT 1996 May 20.
The top one is the X1916-053 optical light curve and the dashed lines
are the expected minimum times from the updated ephemeris given in
eq. (9). The offset between expected and
observed minimum times is due to the phase fluctuation. The bottom
one is the light curve of a nearby comparison star for the identification
of the observation confidence.
\label{tyoplc}}

\figcaption[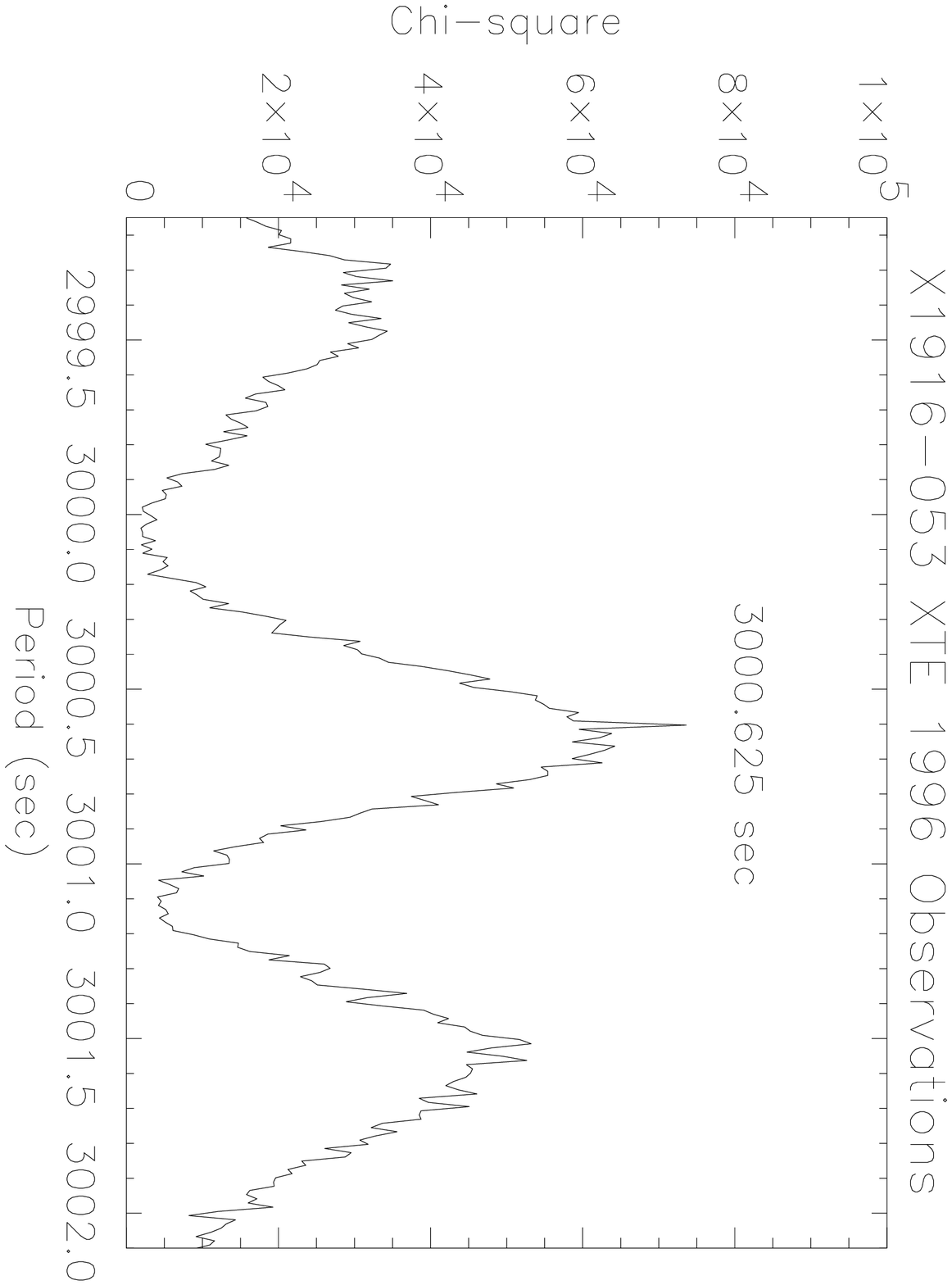]{Result of folding period search for
all RXTE X1916-053 1996 observations.
\label{fesall}}

\figcaption[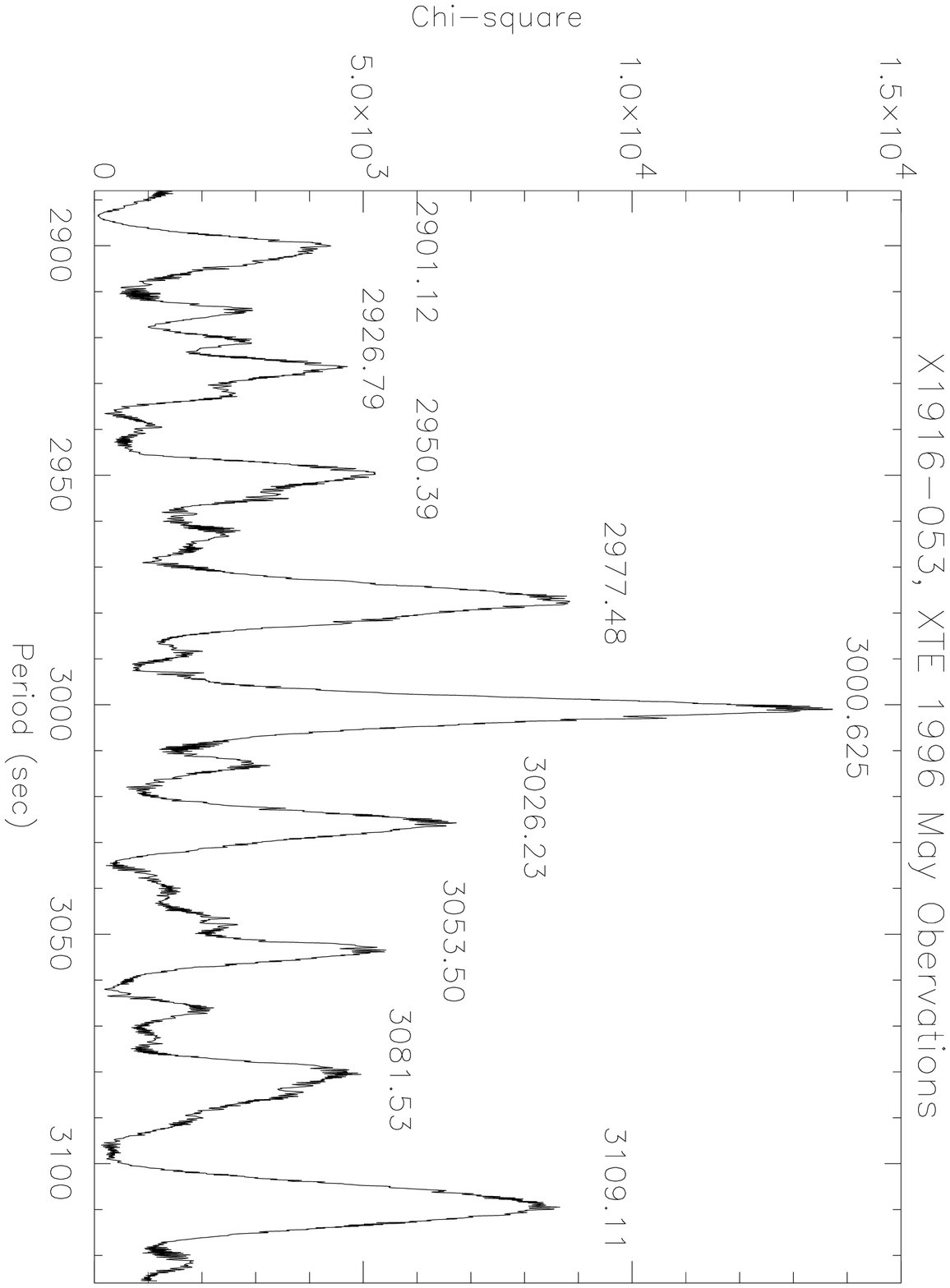]{Result of the folding period search for 
RXTE X1916-053 data during ten
consecutive days in 1996 May.  The 3.9d beat side bands appear
beside the dip period.  The period of first side band on the right of the
3000 sec peak is close to the optical modulation period 3027.551 sec.
\label{fesmay}}

\figcaption[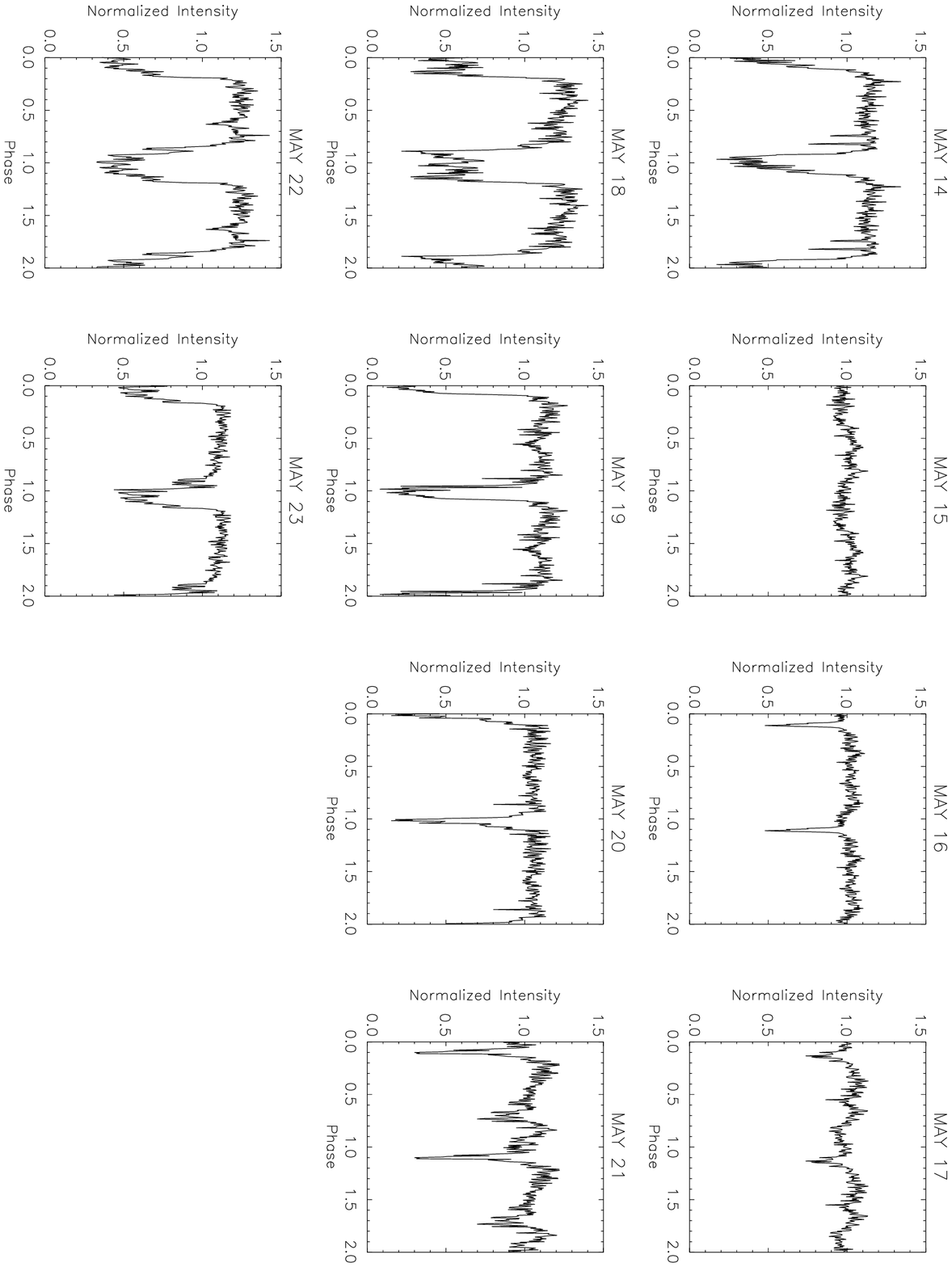]{The folding light 
curves of the 10
consecutive day observation (14-23 May, 1996).  All light curves are folded with a period of
3000.625 sec and an arbitrary but constant phase zero epoch.
\label{tenfef}}

\figcaption[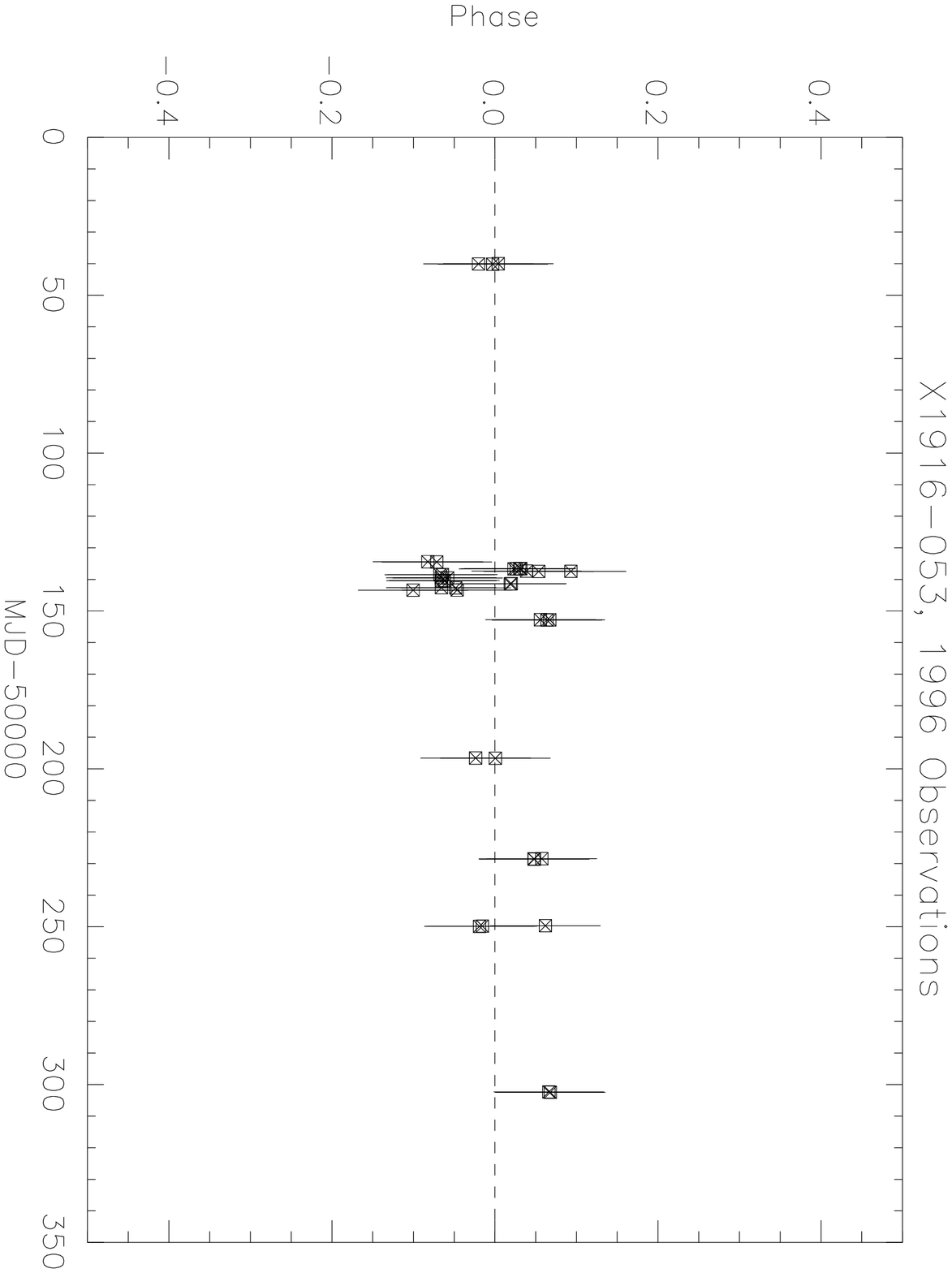]{Dip center time vs. phase of all RXTE 
X1916-053 1996
observations.  The folding period is 3000.625 and the phase zero
epoch is MJD50123.00944.
\label{phase96}}

\figcaption[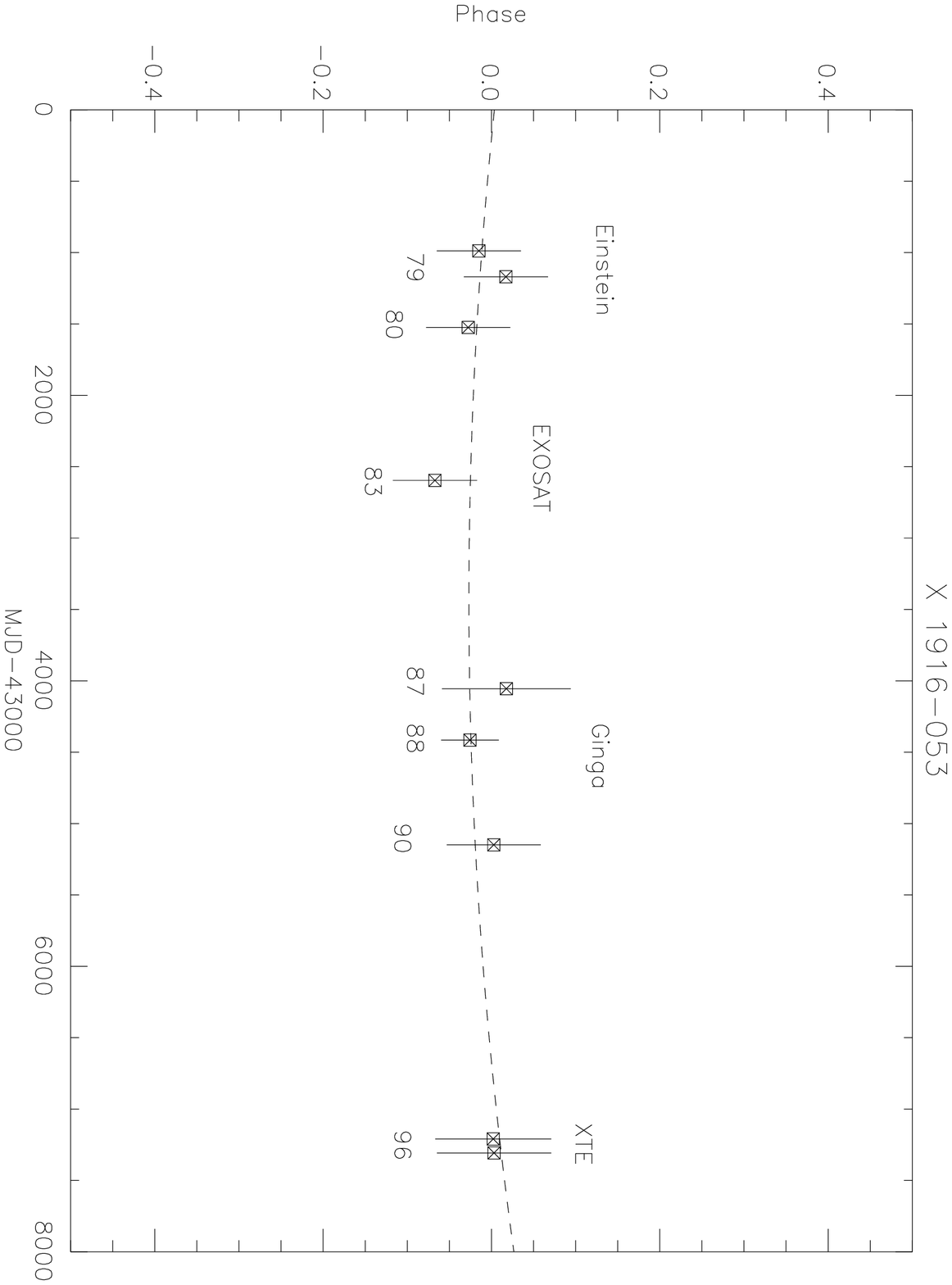]{X-ray dip center time vs. phase 
of X1916-053 1979-1996
observations.  The folding period is 3000.6508 sec and the phase zero
epoch is MJD50123.00944.  The dash line is the quadratic fit result.
\label{phase7996}}

\figcaption[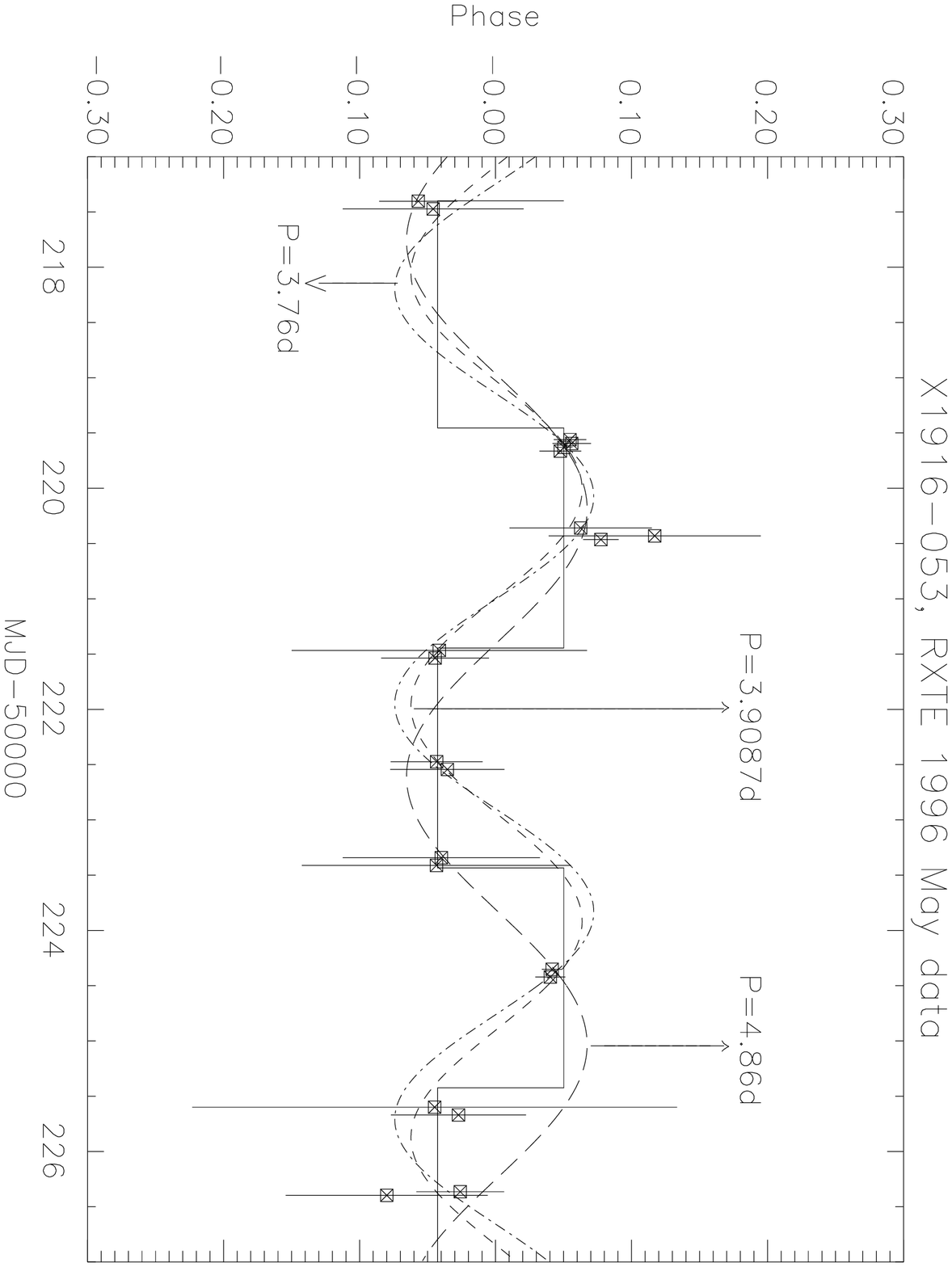]{Phase jitter of the RXTE X1916-053 
1996 May
observation. The solid line is the best square wave fit result
(for P = 3.9087d) and the dashed
lines are the $\chi^2$ minimum fit results (periods labeled) 
for sinusoidal phase delay variations.
\label{jitter}}

\figcaption[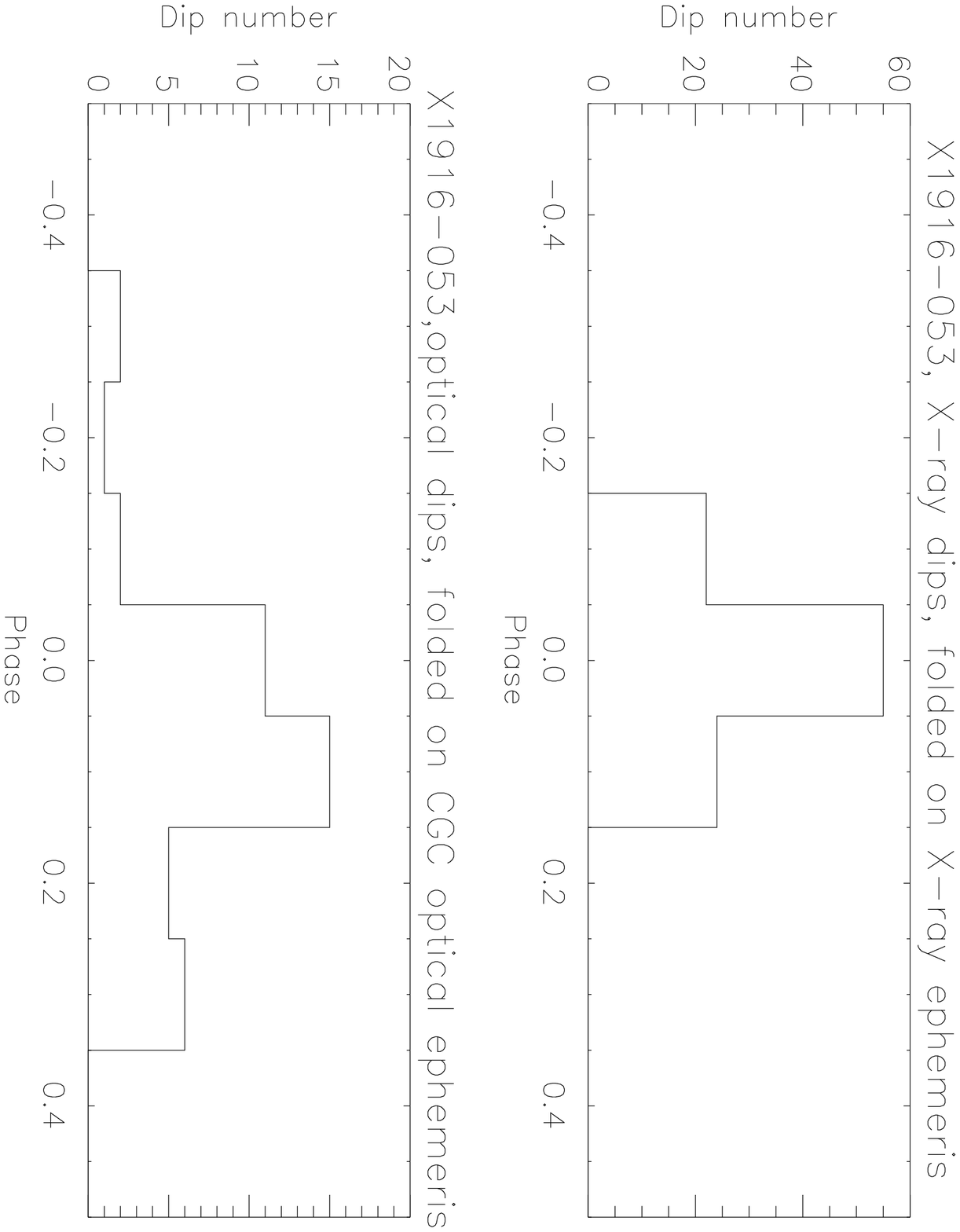]{X1916-053 phase distributions for x-ray
dips (top) and 
optical minima
(bottom) folded by the x-ray and optical ephemerides respectively.  The optical minima show larger phase fluctuations. 
\label{phstat}}

\figcaption[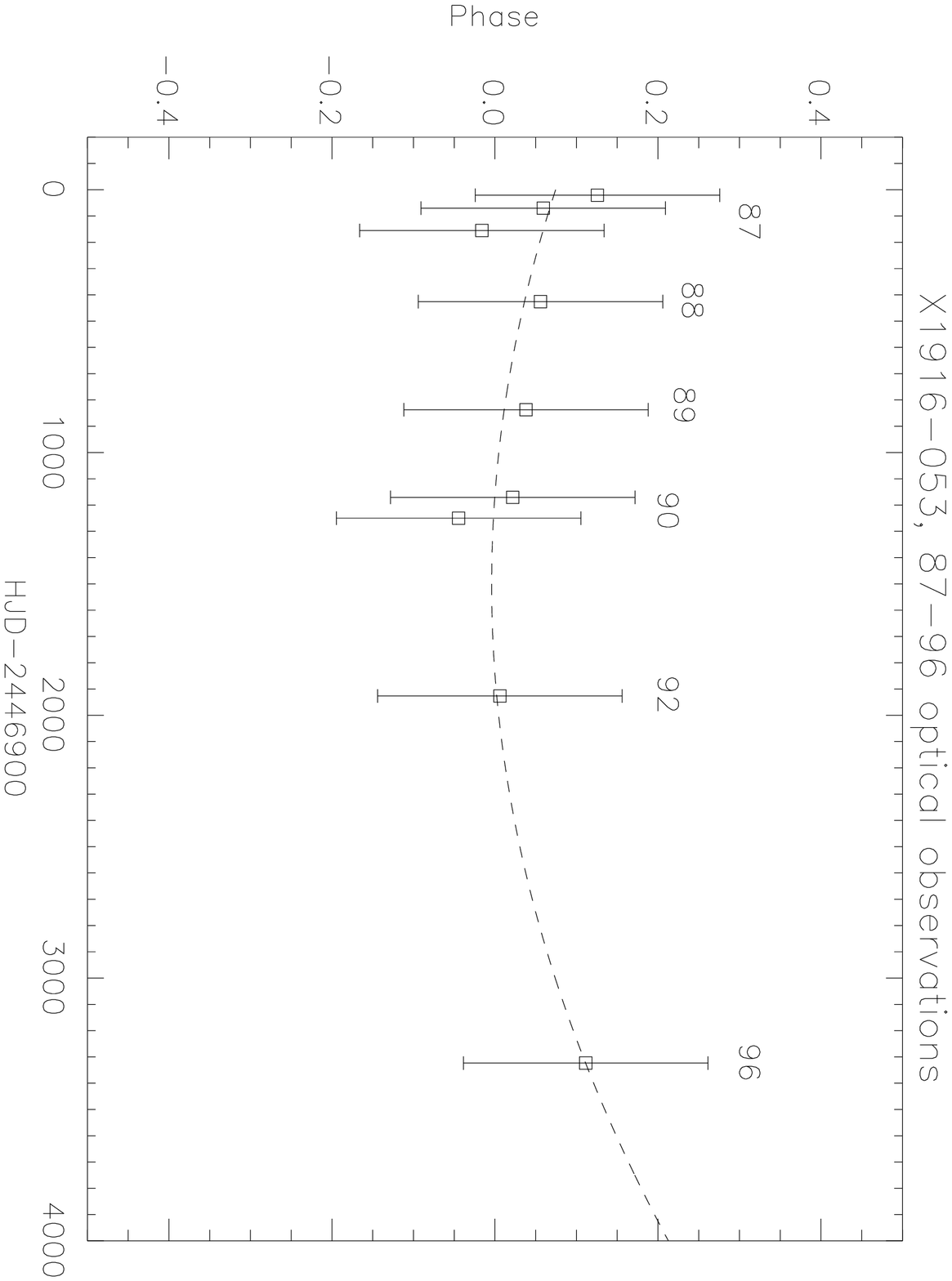]{Optical minimum time vs phase for X1916-
053 '87-'96
observations.  The folding period is 3027.5510 sec and the phase zero
epoch is HJD2446900.0046.  The dashed line is the quadratic fit result,
and the bars show the range of phase jitter for each observation.
\label{qfitop}}

\figcaption[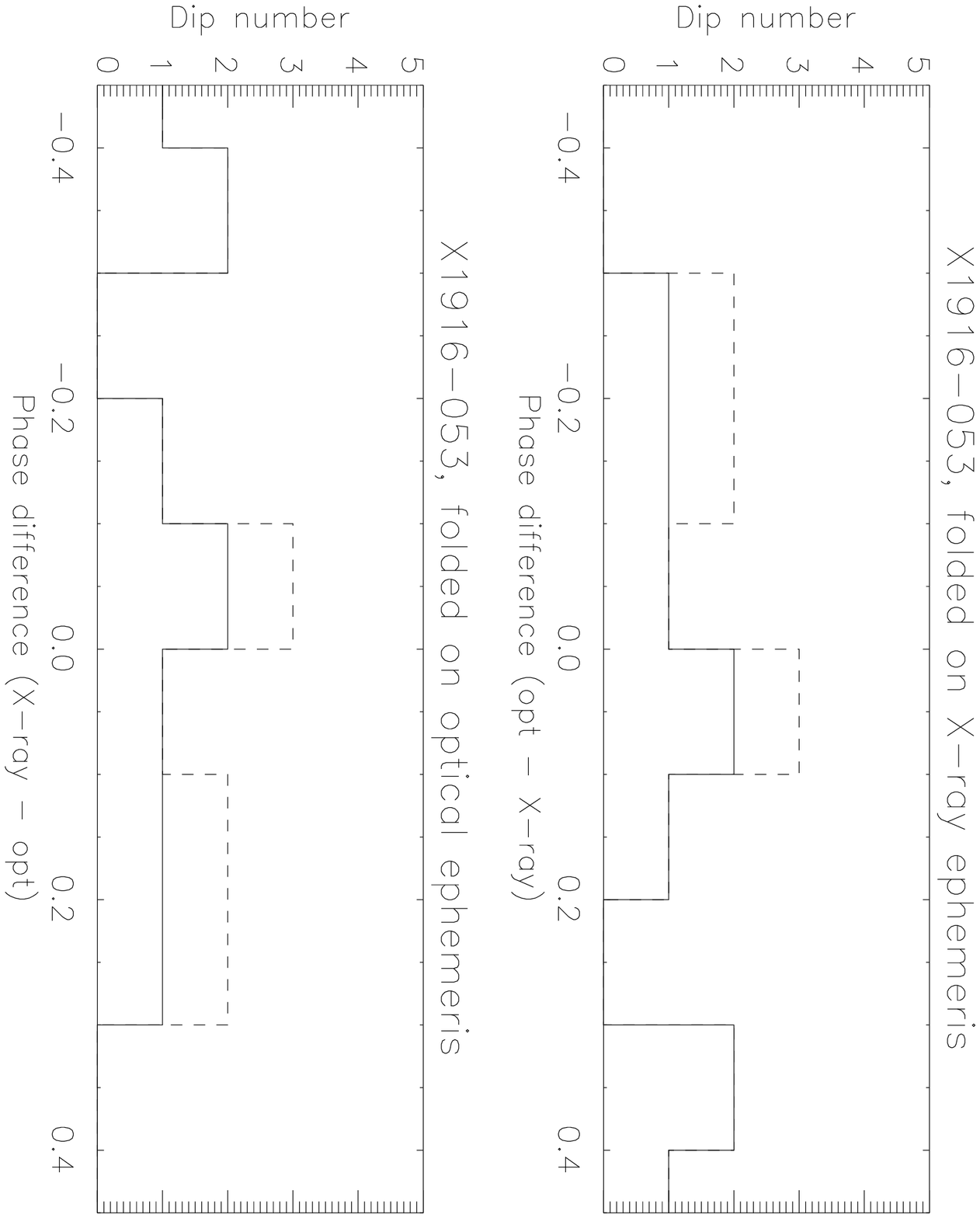]{Phase difference distributions of 12 pair
quasi-simultaneous observed dips/minima on X1916-053.  The plot on the top is the 
distribution of 
$\Delta \phi ( = \phi_{opt} - \phi_x)$ folded by the x-ray ephemeris,
and the
bottom one is the distribution of $\Delta \phi(=\phi_x - \phi_{opt})$ folded by 
the optical ephemeris. The solid lines are the distributions for the 9
pairs of primary dips/minima and dashed lines are for the 3 pairs of secondary dips/minima.
\label{dphstat}}

\figcaption[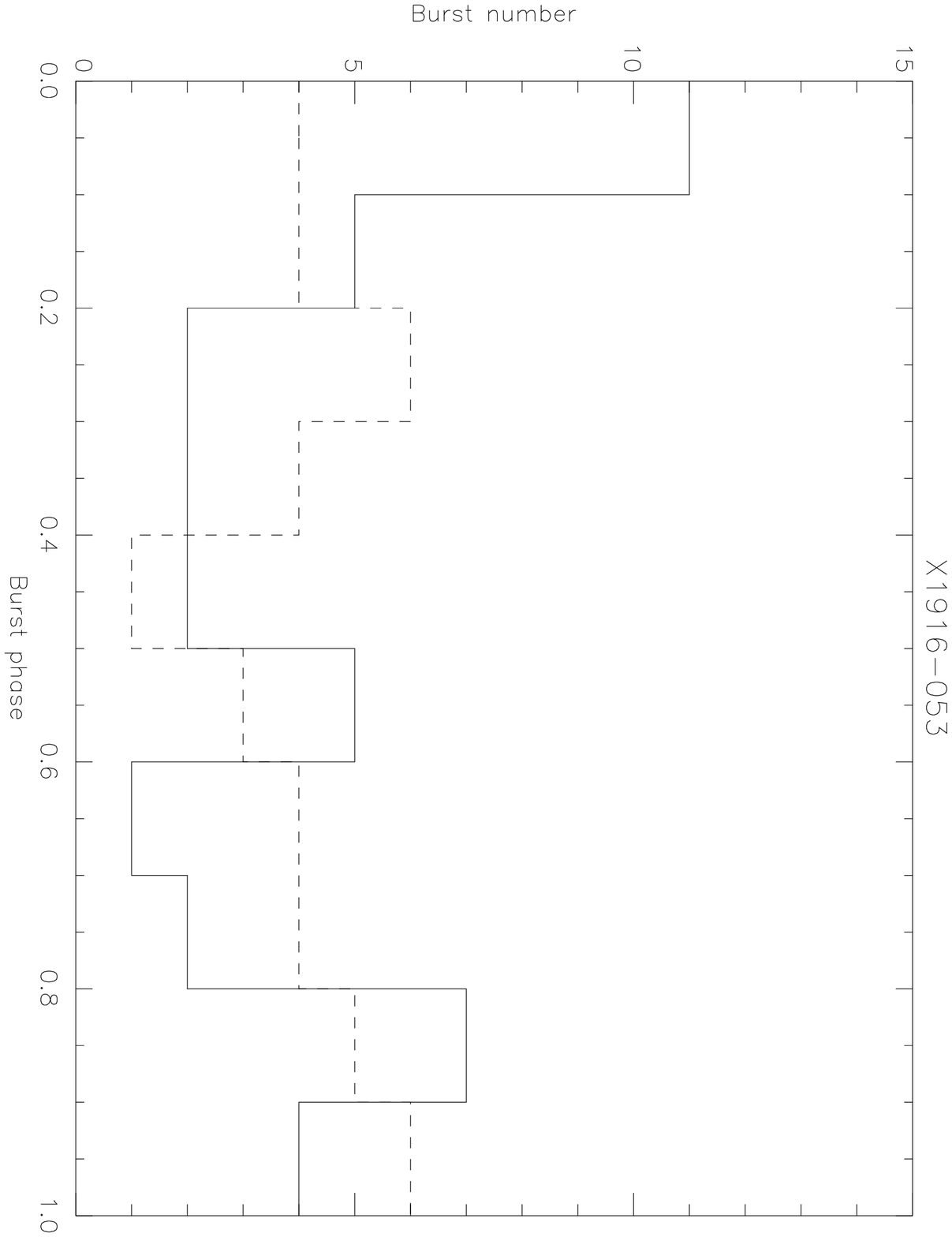]{Phase histogram for 41 x-ray 
bursts (OSO-8 through RXTE data) when folded on the x-ray dip 
ephemeris (primary dips at phase = 0, solid line). The burst arrival times 
appear to be clustered following primary and secondary dips.  The dashed
line is the phase histogram for x-ray burst when folded on the optical
ephemeris. The burst arrival times are independent of the optical modulation.
\label{burststat}}

%tables

\clearpage

\begin{deluxetable}{lcccc}
\tablecaption{Journal of RXTE observations}
\footnotesize
\tablewidth{0pt}
\tablehead{
\colhead{Date} & 
\colhead{Start Time} &
\colhead{Stop time}&
\colhead{Time on Source}& 
\colhead{Mean Band 4}\\
\colhead{dd/mm/yy} &
\colhead{UT(hh:mm:ss)} &
\colhead{UT(hh:mm:ss)} &
\colhead{(sec)}&
\colhead{Count Rate (cts/sec)}
}
\startdata
10/02/96         & 00:14:18   & 05:09:36   &    9808    & 225.4   \\
05/05/96         & 22:16:32   & 03:20:32\tablenotemark{a}  &    7280 &
96.42 \\
14/05/96         & 09:22:08   & 12:55:44   &    8160 & 29.57       \\
15/05/96         & 12:32:16   & 15:11:44   &    6880 & 43.60       \\
16/05/96         & 12:56:48   & 16:31:44   &    8240 & 43.35       \\
17/05/96         & 08:20:16   & 11:49:36   &    7872 & 37.54      \\
18/05/96         & 09:53:52   & 13:20:32   &    7888 & 23.38       \\
19/05/96         & 10:06:40   & 13:34:56   &    7920 & 24.48       \\
20/05/96         & 06:50:24   & 10:16:32   &    7888 & 27.53      \\
21/05/96         & 07:48:00   & 11:22:40   &    8400 & 24.69       \\
22/05/96         & 12:38:08   & 16:26:40   &    8688 & 30.53      \\
23/05/96         & 07:50:56   & 10:23:44   &    6912 & 71.01       \\
01/06/96         & 17:37:04   & 21:36:32   &    9857 & 81.69       \\
15/07/96         & 11:49:52   & 16:09:52   &   10144 & 63.22       \\
16/08/96         & 10:37:52   & 14:55:44   &   10448 & 51.76        \\
06/09/96         & 16:00:32   & 21:26:40   &    9936 & 71.89       \\
29/10/96         & 06:53:36   & 11:22:40   &    9968 & 68.03       \\
\enddata
\tablenotetext{a}{Next day}
\end{deluxetable}

\clearpage

\begin{deluxetable}{rccc}
\tablecaption{Expected vs. observed side bands}
\footnotesize
\tablewidth{0pt}
\tablehead{
\colhead{Harmonic} & 
\colhead{Expected side} &
\colhead{Measured side}&
\colhead{Error of measured} \\
\colhead{index} &
\colhead{band period} &
\colhead{band period} &
\colhead{side band period} \\
\colhead{} &
\colhead{(sec)} &
\colhead{(sec)} &
\colhead{(sec)}
}
\startdata
 1	 &   	   3027.551  &      3026.23    &        3.23        \\	
 2       &	   3055.939  &      3053.50    &        2.93        \\	
 3       &	   3082.826  &      3081.53    &        3.41        \\	
 4       &	   3111.227  &      3109.11    &        4.24        \\	
-1       &	   2974.224  &      2977.48    &        4.01        \\ 	
-2       &	   2948.259  &      2950.39    &        2.85        \\	
-3       &	   2922.743  &      2926.79    &        3.91        \\
-4       &	   2897.665  &      2901.12    &        2.89        \\ 
\enddata
\end{deluxetable}

\clearpage

\begin{deluxetable}{lcccc}
\tablecaption{Measured X1916-053 x-ray dip periods}
\footnotesize
\tablewidth{0pt}
\tablehead{
\colhead{Mission} & 
\colhead{Time of obs} &
\colhead{Period}&
\colhead{Error}&
\colhead{Reference}\\
\colhead{} &
\colhead{(Year)} &
\colhead{(sec)} &
\colhead{(sec)} &
\colhead{}
}
\startdata
OSO-8     &      78        &    3003.6 &     1.8   & 1   \\ 
          &                 &    2995.8 \tablenotemark{a}&     3.6   &
1\\
Einstein  &   79, 80, 81    &    2985   &     10.0   & 2 \\ 
EXOSAT    &    83, 85       &    3015   &     17.0   & 3\\
Ginga     &      87         &    3005.0 &     6.6   & 3\\
	  &                 &    2984.6 \tablenotemark{a}&     6.8   &
3\\
Ginga     &      87         &    3001.2 &     3.0   & 4\\
Ginga     &      89        &    2998.8  &     6.2   & 4\\
Ginga     &      90        &    3000.0  &     1.2   & 4\\
ACSA      &      93         &    3005   &     10    & 5\\
RXTE       &      96        &    3000.625 &    0.020 & 6\\
All        &  79-96         &    3000.6508 &    0.0009 & 6\\

\enddata
\tablenotetext{a}{second candidate} 
\tablerefs{(1) White \& Swank 1982; (2) Walter et al. 1982; (3)
Smale et al. 1989; (4) Yoshida et al. 1995; (5) Church et
al. 1997; (6) this paper}
\end{deluxetable}

%figures
\clearpage

\begin{figure}
\figurenum{\ref{tylc}}
\plotone{band4.lc.May_19.MJD.txt.ps}
\caption{}
\end{figure}

\begin{figure}
\figurenum{\ref{misdip}}
\plotone{band4.lc.May_15.MJD.txt.ps}
\caption{}
\end{figure}

\begin{figure}
\figurenum{\ref{tyoplc}}
\plotone{mag.21.1_av_12to19_comp_7.dipcenter1.ps}
\caption{}
\end{figure}

\begin{figure}
\figurenum{\ref{fesall}}
\plotone{band4.lc.fes.txt.ps}
\caption{}
\end{figure}

\begin{figure}
\figurenum{\ref{fesmay}}
\plotone{band4.lc.May.fes.txt.ps}
\caption{}
\end{figure}

\begin{figure}
\figurenum{\ref{tenfef}}
\plotone{band4.lc.May_tenday.fef.txt.ps}
\caption{}
\end{figure}

\begin{figure}
\figurenum{\ref{phase96}}
\plotone{band4.dip.lifit-2.dat.ps}
\caption{}
\end{figure}

\begin{figure}
\figurenum{\ref{phase7996}}
\plotone{band4.79-96av.fit.p_3000.651.dat.ps}
\caption{}
\end{figure}

\begin{figure}
\figurenum{\ref{jitter}}
\plotone{band4.phasejitter.May.dat.ps}
\caption{}
\end{figure}

\begin{figure}
\figurenum{\ref{phstat}}
\plotone{dipstat_X_opt.ps}
\caption{}
\end{figure}

\begin{figure}
\figurenum{\ref{qfitop}}
\plotone{dipcenterph_av3.dat.ps}
\caption{}
\end{figure}

\begin{figure}
\figurenum{\ref{dphstat}}
\plotone{compX_opt.stat.ps}
\caption{}
\end{figure}

\begin{figure}
\figurenum{\ref{burststat}}
\plotone{1916burstphase_new.stat.ps}
\caption{}
\end{figure}


\begin{references}
\reference{bai87} Bailyn C. D. 1987, ApJ, 317, 737
\reference{blo00} Bloser P. F., Grindlay J. E., Barret D.\& Boirin
L. 2000, ApJ, in press
\reference{bro61} Brouwer, D. \& Clemence, G.D. 1961, Method of
Celestial Mechanics (Academic Press, London)
\reference{cal95} Callanan, P. J., Grindlay. J.E. \& Cool, A. M. 1995,
PASJ, 47, 153 (CGC95)
\reference{chu97} Church, M. J., et al. 1997, ApJ, 491, 388
\reference{chu98} Church, M. J., et al. 1998, A\&A, 338, 556
\reference{coc00} Cocchi, M. et al 2000, 5th Comp. Symp., AIP
Conf. Proc., 510, 203
\reference{fra99} Frank, J., King, A. and Raine, D. 1992,{\it Accretion Power 
in Astrophysics}, Cambridge Univ. Press, p. 99.

\reference{josh88} Grindlay, J. E., Bailyn, C. D., Cohn, H., Lungger, P. M. 
Thorstensen,
J. R. \& Wegner, G. 1988, ApJ, 334, L25
\reference{josh89} Grindlay J. E. 1989, in Proc of the $23^{rd}$ ESLAB Symposium 
on Two
Topics in X-ray Astronomy, J. Hunt,  B. Battrick, ed ESA SP-296, p121
\reference{josh92} Grindlay, J. E. 1992 in Proc. of the 28th Yamada
Conference Frontiers of Astronomy, ed Y. Tanaka, K.Koyama (Tokyo), p69
\reference{hir90} Hirose, M. \& Osaki, Y. 1990, PASJ, 42, 135
\reference{hir91} Hirose, M., Osaki, Y. \& Mineshige, S. 1991, PASJ,
43, 809
\reference{jos81} Joss, P.C. and Li, F.K. 1980, ApJ, 238, 287
\reference{maz97} Mazeh, T. \& Shaham, J. 1979 A\&A, 77, 145
\reference{nar95} Narayan, R. \& Yi, I. 1995 ApJ, 444, 231
\reference{pat93a} Patterson, J., et al. 1993a, PASP, 105, 69 
\reference{pat93b} Patterson, J., et al. 1993b, ApJS, 86, 235
\reference{pat93c} Patterson, J., et al. 1993c, ApJ, 419, 803
\reference{pri84} Priedhorsky, W. C. \& Terrell, J. 1984, ApJ, 208, 661
\reference{pro95} Provencal, J. L., et al. 1995, ApJ, 445, 927
\reference{sch88} Schmidtke, P. C., 1988, A.J., 95, 1528
\reference{ski99} Skillman, D. R., et al. 1999, ASAP, 111, 1281
\reference{sma88} Smale, A.P., Mason, K. O., White, N. E. \& Gottwald,
M. 1988, MNRAS, 232, 674
\reference{sma89} Smale, A.P., Mason, K. O., Williams, O. R. \& Watson,
M. G. 1989 PASJ, 41, 607
\reference{sma92} Smale, A. P., Mukai, K., Williams, O. R., Jones,
M. H. \& Corbet, R. H.,
1992, ApJ, 400, 330
\reference{sol98} Solheim, J. -E., et al. 1998, A\&A, 332, 939
\reference{ste87} Stetson P. 1987, PASP 99, 191
\reference{van93} van der Klis, et al. 1993, MNRAS, 260, 686
\reference{wal82} Walter, F. M., Bowyer, S., Mason, K. O., Clarke, J. T., 
Halpern,
J. \& Grindlay, J. E. 1982, ApJ, 253, L67
\reference{war95} Warner, B. 1995, Cataclysmic Variable Stars (Cambridge University Press)
\reference{whi95} White, N. E. Swank, J. H. 1982 ApJ, 253, L61
\reference{whh88} Whitehurst, R. 1988, MNRAS, 232, 35
\reference{whh89} Whitehurst, R., et al. 1989, Proceedings of 23rd
ESLAB Symposium, vol.1, p127
\reference{yos93} Yoshida, K. 1993, Ph.D. Dissertation, Kanagawa University
\reference{yos95} Yoshida, K., Inoue, H., Mitsuda, K., Dotani \& T Makino, F. 
1995, PASJ,
 47, 141
\reference{zha91} Zhang, E., et al. 1991, ApJ, 381, 534
\end{references}
\end{document}